\documentclass[10pt,notitlepage]{iopart}
\usepackage{a4,epsfig}
\usepackage{url}
\usepackage[dvipsnames]{xcolor}
\usepackage[hidelinks]{hyperref}
\usepackage{lineno}
\usepackage{multirow}
\usepackage{xcolor}
\usepackage{cite}
\usepackage{iopams}
\usepackage[toc,page,title]{appendix}
\usepackage[normalem]{ulem}
\usepackage{graphicx}
\usepackage{subfig}
\begin{document}

\title{Assessing the compact-binary merger candidates reported by the MBTA pipeline in the LIGO-Virgo O3 run: probability of astrophysical origin, classification, and associated uncertainties}

\author{
N.~Andres$^1$, M.~Assiduo$^{2,3}$, F.~Aubin$^{2,3}$, R.~Chierici$^4$, D.~Estevez$^5$\footnote{Corresponding author}, F.~Faedi$^{2,3}$, G.~M.~Guidi$^{2,3}$, V.~Juste$^5$, F.~Marion$^1$, B.~Mours$^5$, E.~Nitoglia$^4$, V.~Sordini$^4$
}

\address{$^1$Laboratoire d'Annecy de Physique des Particules (LAPP), Univ. Grenoble Alpes, Universit\'e Savoie Mont
Blanc, CNRS/IN2P3, F-74941 Annecy, France}
\address{$^2$Universit\`a degli Studi di Urbino 'Carlo Bo,' I-61029 Urbino, Italy}
\address{$^3$INFN, Sezione di Firenze, I-50019 Sesto Fiorentino, Firenze, Italy}
\address{$^4$Institut de Physique des 2 Infinis de Lyon (IP2I) - UMR 5822, Universit\'e de Lyon, Universit\'e Claude
 Bernard, CNRS, F-69622 Villeurbanne, France}
\address{$^5$Universit\'e de Strasbourg, CNRS, IPHC UMR 7178, F-67000 Strasbourg, France}

\begin{abstract}
We describe the method used by the Multi-Band Template Analysis (MBTA) pipeline to compute the probability of astrophysical origin, $p_{astro}$, of compact binary coalescence candidates in LIGO-Virgo data from the third observing run (O3).
The calculation is performed as part of the offline analysis and is used to characterize candidate events, along with their source classification.
The technical details and the implementation are described, as well as the results from the first half of the third observing run (O3a) published in GWTC-2.1. The performance of the method is assessed on injections of simulated gravitational-wave signals in O3a data using a parameterization of $p_{astro}$ as a function of the MBTA combined ranking statistic.
Possible sources of statistical and systematic uncertainties are discussed, and their effect on $p_{astro}$ quantified.
\end{abstract}

\ioptwocol

\section{Introduction} \label{introduction}
As the number of gravitational-wave (GW) detections from compact binary coalescences (CBCs) \cite{GWTC1, GWTC2, GWTC2.1:2021} observed by the Advanced LIGO \cite{LIGO_det:2015} and Advanced Virgo \cite{Virgo_det:2015} detectors increases, the knowledge of GW sources population is improving \cite{O1O2AstroDist:2019, O3aAstroDist:2021}. It allows to compute a probability of astrophysical origin ($p_{astro}$) of GW events and jointly make a source classification based on the the nature of the binary components \cite{Farr:2015, Kapadia:2020}.
The $p_{astro}$ calculation is thus a tool to reveal more events in population-rich areas of the parameter space in complement to the false alarm rate (FAR), which is assigned to a candidate event on the basis of the background behaviour estimated from the data.
The source classification is also a key ingredient to compute merger rates of source-specific compact objects \cite{Fishbach:2018, Mapelli:2018, Vitale:2019, Baibhav:2019} and to inform the population synthesis of such systems \cite{Belczynski:2002, Banerjee:2010, Antonini:2016, Rodriguez:2016, Barrett:2018, O1O2AstroDist:2019, O3aAstroDist:2021}.
Another purpose of the source classification is to provide useful information on GW sources in low-latency searches \cite{LLalerts:2019}. Indeed, as public alerts are sent out to astronomers, the classification can help them decide whether to undertake a follow-up of the source or not.
The ability to provide such information on the nature of the binary components in real-time will thus be of prime importance. For instance, the number of public alerts likely to trigger a search for electromagnetic (EM) and/or high energy neutrinos (HEN) counterparts is expected to reach $\mathcal{O}(1/day)$ in the future O4 observing run \cite{Prospects:2020}.

In this paper, we present the method used offline in O3 to compute the probability of astrophysical origin, $p_{astro}$, and the source classification of CBC search triggers with the Multi-Band Template Analysis (MBTA) pipeline~\cite{MBTA:2016, MBTA:2021}.
MBTA is a matched-filter based search looking first for triggers in each of the two LIGO and Virgo detectors, then a coincidence step is performed, followed by a FAR estimation.
The $p_{astro}$ values are computed as a post-processing task.
MBTA is one of the pipelines developed in the LIGO-Virgo-KAGRA collaboration both for online and offline CBC searches.
It has been used to perform low-latency searches for CBCs ever since the late operation of the first generation of interferometric GW detectors~\cite{LIGO_S6, Virgo_SR, S6_online, O1_BNS_NSBH, O2_online} and started reporting offline results in O3 as part of the archival LIGO-Virgo searches \cite{GWTC2.1:2021, NSBHdiscovery:2021}.

In what follows, we provide a detailed description of how the method developed by  W. Farr, J. Gair, I. Mandel, and C. Cutler, so-called FGMC \cite{Farr:2015}, and extended to a multi-component formalism described in \cite{Kapadia:2020} has been adapted to the MBTA offline analysis for O3, considering three types of CBC sources: Binary Neutron Stars (BNS), Neutron Star-Black hole (NSBH) and Binary Black Holes (BBH). The MBTA results from GWTC-2.1 \cite{GWTC2.1:2021} are then discussed and the performance of the method is assessed via injections of simulated GW signals into the O3a data. A computationally inexpensive method to estimate $p_{astro}$, which is particularly well suited for large injection sets and for use in low-latency searches, is also presented. Finally,
we discuss uncertainties associated with $p_{astro}$ that originate from various assumptions used in the calculation. 

\section{Probability of astrophysical origin and source classification} \label{formalism}
We aim at calculating the probability of astrophysical origin of gravitational wave candidate events that come out of a matched-filter analysis of the detector strain data.
These candidates are referred to as \emph{triggers} throughout the paper. 

We define the astrophysical probability assigned to a trigger as in \cite{RateBBH:2016}. It depends on the ranking statistic $x$ of the trigger defined in \cite{MBTA:2021}, which is a quantity derived from the matched-filter signal-to-noise ratio (SNR). 
In practice we use functions of $x^2$, so that:
\begin{equation}
p_{astro}(x^2) = \frac{\Lambda_1 f(x^2)}{\Lambda_0 b(x^2)+\Lambda_1 f(x^2)}
\label{eq:basicpastro}
\end{equation}
where $\Lambda_0$ (resp. $\Lambda_1$) is the expected number of background (resp. astrophysical) triggers for a given observing time and $b(x^2)~=~p(x^2|\mathrm{noise})$ (resp. $f(x^2)~=~p(x^2|\mathrm{signal})$) is the normalized background (resp. astrophysical) distribution of $x^2$.

\subsection{The multi-component counts posterior}
Since we do not know a priori the expected counts of astrophysical and background triggers, they need to be estimated from the data.
We do so by using a set of $N_{trig}$ triggers with $x$ above a threshold $x_{th}$ low enough to guarantee the sample is dominated by background events.

Using the posterior distribution on the rates given in the FGMC method (equation 21 of \cite{Farr:2015}), which assumes the background and foreground triggers occur as two independent Poisson processes, we define the likelihood of the data given the expected astrophysical and background counts as:
\begin{equation}
p(\vec{x^2}|\Lambda_0,\Lambda_1) \propto e^{-\Lambda_0-\Lambda_1}\prod_{k=1}^{N_{trig}}[\Lambda_0 b(x^2_k) + \Lambda_1 f(x^2_k)]
\end{equation}
where $\vec{x^2}$ is the list of $x^2$ values of the $N_{trig}$ triggers with $x \geq x_{th}$.
An extension of this two-dimensional likelihood to a multi-component likelihood described in \cite{Kapadia:2020} gives:
\begin{equation}
p(\vec{x^2}|\Lambda_0,\vec{\Lambda}_1) \propto e^{-\Lambda_0-\Lambda_1}\prod_{k=1}^{N_{trig}}[\Lambda_0 b(x^2_k) + \vec{\Lambda}_1\cdot \vec{f}(x^2_k)]
\label{eq:multicomplikeli}
\end{equation}
with $\vec{\Lambda}_1$=$ \{\Lambda_{\alpha}\}$ a vector of expected counts for the three categories we consider in this paper $\alpha\in\{\textsc{BNS},\textsc{NSBH},\textsc{BBH}\}$, $\Lambda_1$=$ \sum_{\alpha} \Lambda_{\alpha}$ and $\vec{f}(x^2)$=$\{f_{\alpha}(x^2)\}$ a vector of foreground distributions for the three types of sources.
Bayes' theorem gives the multi-component posterior distribution of counts:
\begin{equation}
p(\Lambda_0,\vec{\Lambda}_1|\vec{x^2}) \propto \pi(\Lambda_0, \vec{\Lambda}_1)~p(\vec{x^2}|\Lambda_0,\vec{\Lambda}_1)
\end{equation}
with $\pi(\Lambda_0, \vec{\Lambda}_1)$ the prior distribution on the background and astrophysical counts.

In practice, since we select $\mathcal{O}(10^4)$ triggers which are background dominated, we approximate $\Lambda_{0}$ by $N_{trig}$. For the astrophysical categories we choose a uniform prior on counts for BNS and NSBH, given that no events were detected by MBTA in these categories during O3a \cite{O1_BNS_NSBH}, and a Poisson-Jeffreys prior on counts for BBH \cite{Supplement:2016}.

\subsection{Dividing the search parameter space}
\label{sec:binning}
The matched filter technique performed by the MBTA pipeline to search for GW signals uses a bank of templates covering a redshifted component mass space from $1~M_{\odot}$ to $195~M_{\odot}$ with total mass $\leq200~M_{\odot}$.
Since different CBC source categories are expected to be detected in particular areas of the parameter space, we choose to divide the parameter space in several parts that we call \emph{bins}, to be able to track the foreground distributions accordingly.
As far as the background distributions are concerned, the binning of the parameter space also allows us to better monitor the background variations.

The parameters chosen to segment the parameter space are the detector-frame chirp mass $\mathcal{M}~=~(m_1 m_2)^{3/5}/(m_1+m_2)^{1/5}$ and the mass ratio $q=m_2/m_1$ with $m_1$ and $m_2$ the redshifted masses of the binary components and $m_1 \geq m_2$.

The $`` \mathcal{M} - q"$ space is divided into $45$ bins in $\mathcal{M}$ and $4$ bins in $q$ leading to $N_{bins}=165$ bins containing templates, with the number of templates per bin varying from $1$ to about $23000$ (median: 1075 and mean: 4412).
The fine segmentation of the parameter space in the dimension of the detected chirp mass is motivated by the fact that it is a well measured parameter in GW signals.
On the other hand, the mass ratio is poorly measured and only a coarse division of the parameter space is performed.
The background and foreground estimates in the resulting bins are described in sections \ref{backg} and \ref{foreg}.

\subsection{Multi-detector framework}
\label{multidet}
The coincident triggers of the search can occur in different detector combinations, which results in $N_{coinc}=7$ coincidence types. They are labeled as HL, HV, LV for double coincidences during double-detector time, HL-Von, HV-Lon, LV-Hon for double coincidences in triple-detector time and HLV for triple coincidences, where H, L and V stands respectively for LIGO-Hanford, LIGO-Livingston and Virgo.
Due to the heterogeneous sensitivities of the detectors and their orientations, the different coincidence types do not have the same ability to capture astrophysical signals.
Thus, to account for this effect, we use weights applied to the foreground distributions (see section \ref{foreg}).
They are computed with injections of simulated GW signals from table~\ref{tab:param_injections} performed in O3a data and counting the fraction of recovered injections during triple-detector time. The corresponding weights $c_j$ are $c_{\mathrm{HL-Von}}=0.86$, $c_{\mathrm{HV-Lon}}=0.01$, $c_{\mathrm{LV-Hon}}~=~0.02$, $c_{\mathrm{HLV}} = 0.11$.
However, the three detectors are not always in observation mode at the same time and double coincidences can occur when the third detector is off. We thus construct the corresponding weights by adding the double coincidence weights to the triple coincidence weight.
For instance, the HL weight is computed as $c_{\mathrm{HL}} = c_{\mathrm{HL-Von}} + c_{\mathrm{HLV}}$.

\subsection{$p_{astro}$ and classification}
Taking the definition of $p_{astro}$ from equation \ref{eq:basicpastro} and marginalizing over the posterior distribution of expected astrophysical and background counts, one can define the probability for a trigger with combined ranking statistic $x$ of belonging to a source category $\alpha$:
\begin{equation}
p_{\alpha}(x^2) = \int_0^{\infty} p(\Lambda_0,\vec{\Lambda}_1|\vec{x^2}) \frac{\Lambda_{\alpha} f_{\alpha}(x^2)}{\Lambda_0 b(x^2)+\vec{\Lambda}_1 \cdot \vec{f}(x^2)} \mathrm{d}\Lambda_0 \mathrm{d}\vec{\Lambda}_1
\label{eq:margpastro}
\end{equation}
We note that this equation depends on the type of coincidence of the trigger and on its corresponding bin of the parameter space. 

This step in the analysis can be time consuming, depending on the number of triggers analyzed and the number of categories considered.
We use a Markov Chain Monte Carlo integration over the three astrophysical counts and approximate $\Lambda_0\sim N_{trig}$.

The probability of astrophysical origin is in turn:
\begin{equation}
p_{astro}(x^2)=\sum_{\alpha} p_{\alpha}(x^2)
\end{equation}
By definition, the probability of a trigger to be of background origin is $p_0(x^2) = 1 - p_{astro}(x^2)$.

In the next sections we describe how $f_{\alpha}$ and $b$ depend on the bin and coincidence type.
\section{Background density distributions} \label{backg}
The background density distributions are computed for the various types of coincidence using a similar process to MBTA's FAR computation \cite{MBTA:2021}. It depends on the single-detector SNR thresholds and it is built by making fake coincidences of all possible combinations of single-detector triggers with identical templates from different detectors over a period of time.
This is done by removing significant coincident GW detections from the data.

However, the background used in $p_{astro}$ is different from the background computed in the standard search to assign a FAR value to candidate events.
Indeed, the background for $p_{astro}$ is computed individually for the $N_{bins}$ bins in the ``$\mathcal{M}-q$" space, whereas the standard search uses a coarse division of the template bank into three regions in the ``$m_1-m_2$" space.
Another difference is the timespan over which the background is estimated. 
The standard search uses periods of about one week over O3, whereas the $p_{astro}$ calculation uses a background estimated over periods of a few months.
Moreover, since the tails of the background distributions may be spoiled by remaining single-detector triggers from astrophysical events, the background for $p_{astro}$ only considers data when at least the Hanford and Livingston detectors are observing (i.e. HL+HLV time).
This has the benefit of removing most of the single detector events from the background distributions, since the majority of the detected GW events are of HL(-Von) type.

The background rate distributions are noted as $\hat{b}_{i,j}(x^2)$, where $i$ and $j$ hereafter reflect the dependence on the bins and the coincidence types, respectively.
They satisfy the following relation:
\begin{equation}
\sum_{i=1}^{N_{bins}}\sum_{j=1}^{N_{coinc}} \int_{0}^{T_j} \int_{x^2_{th}}^{\infty}  \hat{b}_{i,j}(x^2)~\mathrm{d} t~\mathrm{d} x^2 = \Lambda_0
\end{equation}
where $T_j$ is the observing time for coincidence type~$j$.

The background density distribution in bin $i$ and for coincidence type $j$ is thus:
\begin{equation}
b_{i,j}(x^2) = \frac{T_j}{\Lambda_0}~\hat{b}_{i,j}(x^2)
\end{equation}
We note that the background rates for double coincidences is independent of the state of the third detector, such that, e.g., $\hat{b}_{i,\mathrm{HL}}(x^2) = \hat{b}_{i,\mathrm{HL-Von}}(x^2)$.
\section{Foreground density distributions} \label{foreg}
The foreground density distributions are estimated by assuming that the number of sources detected above a SNR threshold is proportional to the detection volume, which is assumed to be constant over the observing run.
In the current implementation, those distributions do not depend on the specific source population \cite{Schutz:2011}, and they are independent of the redshift, since most of the binary mergers observed during O3 occured at low redshift ($z<1$).
These assumptions will be revised when the detectors sensitivity allows to reach sources with greater luminosity distances or the population evolution with redshift is better measured.

The ranking statistic $x$ is defined in MBTA as the SNR rescaled by a factor that downgrades suspected background events (see section 4 of \cite{MBTA:2021}).
This rescaling factor is usually $1$ for astrophysical events and therefore $x$ is distributed as the SNR.
This implies that the cumulative foreground distribution $F(x)$ above a given threshold $x_{th}$ scales as $x^{-3}$.
It is then straightforward to derive the normalized foreground density distribution as a function of $x^2$:
\begin{equation}
\tilde{f}(x^2) = \frac{3}{2} \frac{(x_{th}^2)^{1.5}}{(x^2)^{2.5}} 
\end{equation}

\subsection{Bin weights and population models}
\label{subregionweights}
Since we want to describe the distribution over our bins of the three CBC astrophysical categories, we further consider source-specific, bin-specific foreground density distributions:
\begin{equation}
f_{\alpha,i}(x^2)=w_{\alpha,i}~\tilde{f}(x^2)
\end{equation}
where $w_{\alpha,i}$ represents the relative abundance of type-$\alpha$ sources detected in bin $i$.
This has been estimated by performing injections of simulated GW signals into LIGO-Virgo O3a data.
We set by definition:
\begin{equation}
w_{\alpha,i} = \frac{N_{\alpha,i}}{\sum_{k=1}^{N_{bins}} N_{\alpha,k}}
\end{equation}
where $N_{\alpha,i}$ is the number of recovered injections of source type $\alpha$ with $x \geq x_{th}$ inside bin~$i$.
The injection sets used are the same as the ones described in \textsc{GWTC-2.1} to compute the source hypervolumes probed by the search \cite{GWTC2.1:2021}.
The parameter distributions of the injections are given in table \ref{tab:param_injections}.

\begin{table*}
\centering
\begin{tabular}{cccccccc}
\hline
\multicolumn{2}{c}{ } & Mass & Mass &  Spin & Spin & Redshift & Maximum  \\
\multicolumn{2}{c}{ } & distribution & range ($M_{\odot}$) & range & orientations & evolution & redshift  \\
\hline
\rule{0pt}{2.6ex}
\multirow{8}{*}{} & \multirow{2}{*}{BBH (inj)} &  $p(m_1) \propto m_1^{-2.35}$ & $2<m_1<100$ & \multirow{2}{*}{$\left|\chi_{1,2}\right|<0.998$} & \multirow{2}{*}{isotropic} &\multirow{2}{*}{$\kappa = 1$} & \multirow{2}{*}{$1.9$} \\
 & & $p(m_2|m_1) \propto m_2$ & $2<m_2<100$ & & & & \\[0.05\normalbaselineskip]
 & \multirow{2}{*}{{BBH (pop)}} &  \multirow{2}{*}{\textsc{Power Law + Peak}} & $5<m_1<80$ & \multirow{2}{*}{$\left|\chi_{1,2}\right|<0.998$} & \multirow{2}{*}{isotropic} & \multirow{2}{*}{$\kappa = 0$} & \multirow{2}{*}{$1.9$} \\ 
 & &  & $5<m_2<80$ &  & &   \\
 & \multirow{2}{*}{NSBH} & $p(m_1) \propto m_1^{-2.35}$ & $2.5<m_1<60$ & $\left|\chi_1\right| < 0.998$ & \multirow{2}{*}{isotropic} &  \multirow{2}{*}{$\kappa = 0$} &  \multirow{2}{*}{$0.25$} \\
 & & uniform & $1<m_2<2.5$ & $\left|\chi_2\right| < 0.4$ & & &  \\[0.05\normalbaselineskip]
 & \multirow{2}{*}{BNS} & \multirow{2}{*}{uniform} & $1<m_1<2.5$ & \multirow{2}{*}{$\left|\chi_{1,2}\right| < 0.4$} & \multirow{2}{*}{isotropic} &  \multirow{2}{*}{$\kappa = 0$} & \multirow{2}{*}{$0.15$}  \\
 & & & $1<m_2<2.5$ & & & & \\
\hline
\end{tabular}
\caption{Parameter distributions used to generate injections of GW signals. The BBH fiducial population used to reweight the BBH injections is also provided. The redshift evolution defines the distribution of signals over redshift such as $p(z)\propto (1+z)^{\kappa-1} dV_c/dz$ with $V_c$ the comoving volume. \label{tab:param_injections}}
\end{table*}

Astrophysical population models are key ingredients in $p_{astro}$ calculations since they are used to build the foreground distributions.
For BNS and NSBH we use the models from the injections.
For the BBH injections, we use a population reweighting based on recent astrophysical observations using the \textsc{Power~Law~+~Peak} model defined in Appendix B.2 of \cite{O3aAstroDist:2021} and we assume a non-evolving comoving merger rate density with redshift.
The parameters used for our fiducial BBH population are the same as the ones used in \cite{GWTC2.1:2021}; $\alpha=2.5$, $\beta=1.5$, $m_{min}~=~5~M_{\odot}$, $m_{max}~=~80~M_{\odot}$, $\lambda_{peak}=0.1$, $ \mu_{m}=34~M_{\odot}$, $\sigma_{m}~=~5~M_{\odot}$, $\delta_{m}=3.5~M_{\odot}$ and $\kappa=0$.

\begin{figure} [h]
\centerline{\includegraphics[width=0.5\textwidth]{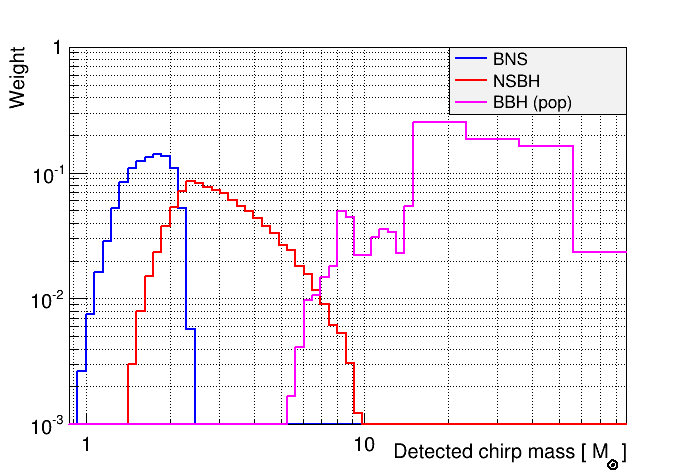}}
\caption{Normalized distributions of the detected chirp mass for the recovered BNS, NSBH and BBH injections. The y-axis gives the bin weights ($w_{\alpha,i}$) marginalized over the mass ratio.\label{fig:foregroundweights}}
\end{figure}

The weights $w_{\alpha,i}$ to construct the foreground density distributions are then directly extracted from the distribution of the detected injections in the ``$\mathcal{M}-q$'' space.
The normalized distributions of detected chirp masses for the three source categories of CBC injections are shown in figure \ref{fig:foregroundweights}.
As expected from the assumed mass gap between the BNS and BBH populations from $2.5~M_{\odot}$ to $5~M_{\odot}$, the simulated BNS and BBH signals are recovered in distinct bins of the parameter space. However, the simulated NSBH signals exhibit a significant overlap with the other two categories.

\subsection{Multi-detector weights}
As stated earlier, we apply weights derived from the $c_j$, defined in section~\ref{multidet}, on the foreground density distributions depending on the coincidence type $j$ of the selected triggers.

Let's consider $\Lambda_{\alpha,i,j}$, the expected number of events of category $\alpha$ inside a bin $i$ for a given type of coincidence~$j$:
\begin{equation}
\Lambda_{\alpha,i,j} = \Lambda_{\alpha}~w_{\alpha,i}~w_j
\end{equation}
with
\begin{equation}
w_j = \frac{T_j c_j}{\sum_{k=1}^{N_{coinc}} T_k c_k}
\label{eq:newweights}
\end{equation}
It satisfies:
\begin{equation}
\Lambda_{\alpha} = \sum_{i=1}^{N_{bins}} \sum_{j=1}^{N_{coinc}} \Lambda_{\alpha,i,j}
\end{equation}

Finally, the foreground distribution for coincidence type $j$ is expressed as:
\begin{equation}
f_{\alpha,i,j}(x^2)=w_{\alpha,i}~w_{j}~\tilde{f}(x^2)
\end{equation}
with the foreground distributions satisfying:
\begin{equation}
\sum_{i=1}^{N_{bins}}\sum_{j=1}^{N_{coinc}} \int_{x^2_{th}}^{\infty} f_{\alpha,i,j}(x^2)~\mathrm{d} x^2 = 1
\end{equation}

In practice, for the O3 catalog papers \cite{GWTC2.1:2021, GWTC3:2021}, these weights were not computed in this way but rather as:
\begin{equation}
w_j = \frac{c_j}{\sum_{k=1}^{N_{coinc}} c_k}
\end{equation}
which involves a crude approximation that the $T_j$ were all the same.
The effect of this approximation on $p_{astro}$ is discussed in section \ref{uncertainties}.

\section{MBTA results in \textsc{GWTC-2.1}} \label{catalogresults}
\begin{figure*}[h]
  \centering
  \subfloat[]{\includegraphics[width=0.45\textwidth]{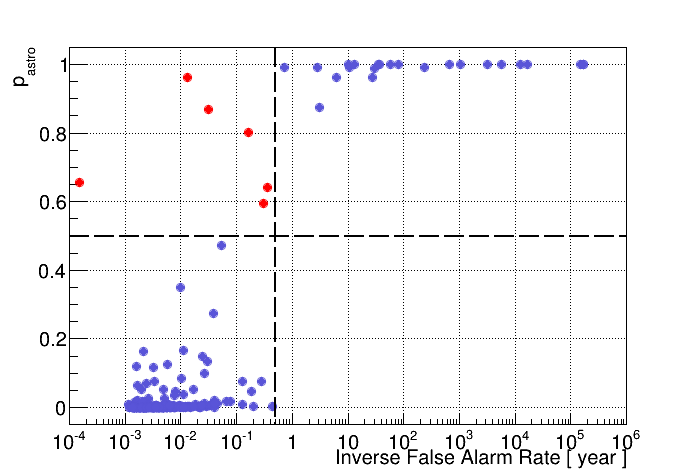}\label{fig:O3apastroifar}}
  \hfill
  \subfloat[]{\includegraphics[width=0.45\textwidth]{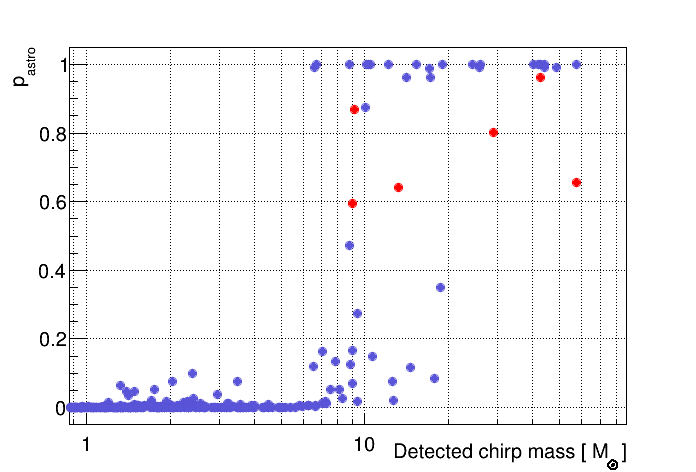}\label{fig:O3apastromchirp}}
  \caption{$p_{astro}$ as a function of the IFAR (a) or the detected chirp mass (b) for the O3a MBTA candidate events with either $\mathrm{IFAR}<0.5~day$ or $p_{astro}>0.5$. The dashed vertical (resp. horizontal) line shows the IFAR (resp. $p_{astro}$) value of $0.5~yr$ (resp. 0.5). The red dots correspond to O3a events recovered with $p_{astro} > 0.5$ but with $\mathrm{IFAR} < 0.5~yr$.}
\end{figure*}

In \textsc{GWTC-2.1}, the MBTA pipeline provided source classification and probability of astrophysical origin for candidate events using the method described in the above sections. All the significant candidates found are of the BBH type.
In figure \ref{fig:O3apastroifar} we show the $p_{astro}$ values of the MBTA O3a candidate events with a $\mathrm{FAR} < 2/day$ or $p_{astro}~>~0.5$ as a function of their inverse false alarm rate (IFAR).
We also draw lines corresponding to two thresholds used in GWTC-2~\cite{GWTC2} (IFAR $= 0.5~yr$) and in GWTC-2.1 \cite{GWTC2.1:2021} ($p_{astro} =~0.5$) used to report full parameter estimation on the candidate events.

The candidate events with $\mathrm{IFAR}~>~0.5~yr$ have $p_{astro}>0.5$ and most of the ones with $\mathrm{IFAR}~<~0.5~yr$ have $p_{astro}<0.5$.
However, six candidate events have an $\mathrm{IFAR}~<~0.5~yr$ but $p_{astro}~>~0.5$.
Several effects explain this difference.
First, given the observed BBH rate, a $p_{astro}$ cut of 0.5 does not translate into a FAR cut of $2/yr$.
Furthermore, the binning of the search parameter space used for $p_{astro}$ plays at least two important roles.
It enables to capture the variations of the assumed BBH population across the parameter space, revealing candidates in population-rich areas.
Here, the foreground density distribution for BBH signals favors high-mass systems as shown in figure~\ref{fig:foregroundweights}.
Moreover, it allows to better track the background variations across the parameter space, especially in the high-mass region of the parameter space where the density of templates is low.
This is different from how the current MBTA standard search computes the FAR of triggers on only three search regions \cite{MBTA:2021}, which can result in assigning a pessimistic FAR value to high-mass triggers with low SNR.
This effect can be seen in figure \ref{fig:O3apastromchirp}, where the two candidate events with the highest FARs ($\mathrm{IFAR} < 1~week$) and $p_{astro} > 0.5$ occurred at high detected chirp mass ($\mathcal{M} > 40~M_{\odot}$).

\section{$p_{astro}$ parameterization} \label{pastroparam}
The above method is well suited for offline analysis on real data but the computational cost of the $p_{astro}$ marginalization makes it difficult to use in an online search.
Moreover, when dealing with large sets of injections, we want to be able to assign $p_{astro}$ values using the observed merger rates (not counting the injections as part of the detections).
We thus extract a simple parameterization of $p_{astro}$ as a function of $x^2$, which can be used both for injections or during online searches.
The method assumes that the number of terrestrial and astrophysical expected counts are known for a given observed space-time volume.

\begin{figure} [h]
\centerline{\includegraphics[width=0.5\textwidth]{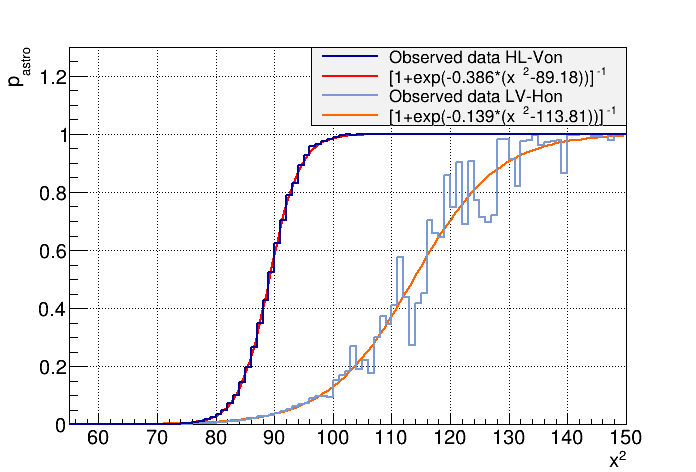}}
\caption{$p_{astro}$ as a function of the combined ranking statistic squared for the HL-Von (LV-Hon) coincidence type with chirp mass and mass ratio: $1.15~ M_{\odot} \leq \mathcal{M}<1.23~ M_{\odot}$ and $0.50~\leq ~q~<0.75$ ($12.13~ M_{\odot} \leq \mathcal{M}<13.00~ M_{\odot}$ and $0<q<0.25$). The dark (light) blue lines are the observed data computed with equation \ref{eq:pastroparam} and the red (orange) curve is the fit with the parameterization proposed in equation~\ref{eq:fitparam}.
\label{fig:O3apastroparambins}}
\end{figure}

To compute $p_{astro}$ values as a function of $x^2$ based on the observations, we take the mean value of the MBTA O3a posterior distributions of counts for the three astrophysical categories $\langle \vec{\Lambda}_1 \rangle$ and we rewrite the astrophysical probability for each bin $i$ and coincidence type $j$ as:
\begin{equation}
p_{astro,i,j}(x^2) = \frac{\langle \vec{\Lambda}_1\rangle \cdot \vec{f}_{i,j}(x^2)}{\Lambda_0 b_{i,j}(x^2) + \langle \vec{\Lambda}_1\rangle \cdot \vec{f}_{i,j}(x^2)}
\label{eq:pastroparam}
\end{equation}
Then we use the following parameterization to fit the data:
\begin{equation}
p_{astro,i,j}(x^2) = \Big[ 1 + \exp\Big(-a_{i,j} \cdot (x^2 - x^2_{50\%,i,j})\Big) \Big]^{-1}
\label{eq:fitparam}
\end{equation}
where $x_{50\%,i,j}$ is the combined ranking statistics at $p_{astro,i,j}=0.5$ and $a_{i,j}$ is the slope of the parameterized curve at $x^2_{50\%,i,j}$.
Figure \ref{fig:O3apastroparambins} shows examples of the $p_{astro}$ parameterization in one HL-Von bin, typical case for O3a candidate events, and in one LV-Hon bin, typical case of a noisier bin for an unlikely type of coincidence.

One can see that another advantage of this method is that it allows to interpolate $p_{astro}$ values around $x_{50\%}$ and thus gives a better precision, compared to using discrete values in the region where $p_{astro}$ varies quickly with $x^2$.
Also, since the background is estimated empirically using fake coincidences of detector triggers, the background distribution can be limited to fairly low values of $x^2$ inside low template density bins. The parameterization allows to extrapolate the $p_{astro}$ trend to higher combined ranking statistic for bins where $p_{astro}$ cannot formally be estimated anymore.
It also helps to smooth out statistical fluctuations of the background that can occur in low statistics bins, as illustrated on figure~\ref{fig:O3apastroparambins}.

\begin{figure} [h]
\centerline{\includegraphics[width=0.46\textwidth]{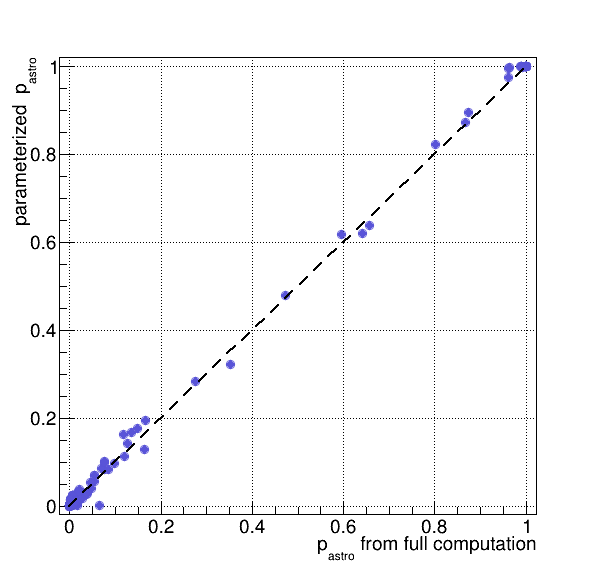}}
\caption{$p_{astro}$ values computed using the parameterization as a function of the $p_{astro}$ values marginalized over the multi-component counts posterior for the O3a MBTA candidate events. The dashed diagonal is the first bisector.\label{fig:O3adataparam}}
\end{figure}

To illustrate the consistency of the parameterized $p_{astro}$ values with those produced by the full computation, we compare them for the O3a GW~candidates reported by MBTA in GWTC-2.1.
Figure~\ref{fig:O3adataparam} shows that they are in good agreement.
The differences between the values may arise from at least three features.
The first one is the use of the mean value of the posterior distributions of astrophysical counts in the parameterization instead of computing marginalized $p_{astro}$ values over these distributions for the full computation.
The second one is that the parameterization interpolates $p_{astro}$ as a function of $x^2$, which is interesting, especially around $p_{astro}=0.5$ where the absolute difference between two consequent $p_{astro}$ values is on average $0.06$ for the BBH bins.
The third one is linked to the goodness of the fit which means that the parameterization can give different $p_{astro}$ values than the full method if it does not match exactly the data points.

\begin{figure*} [h]
\centerline{\includegraphics[width=1.05\textwidth]{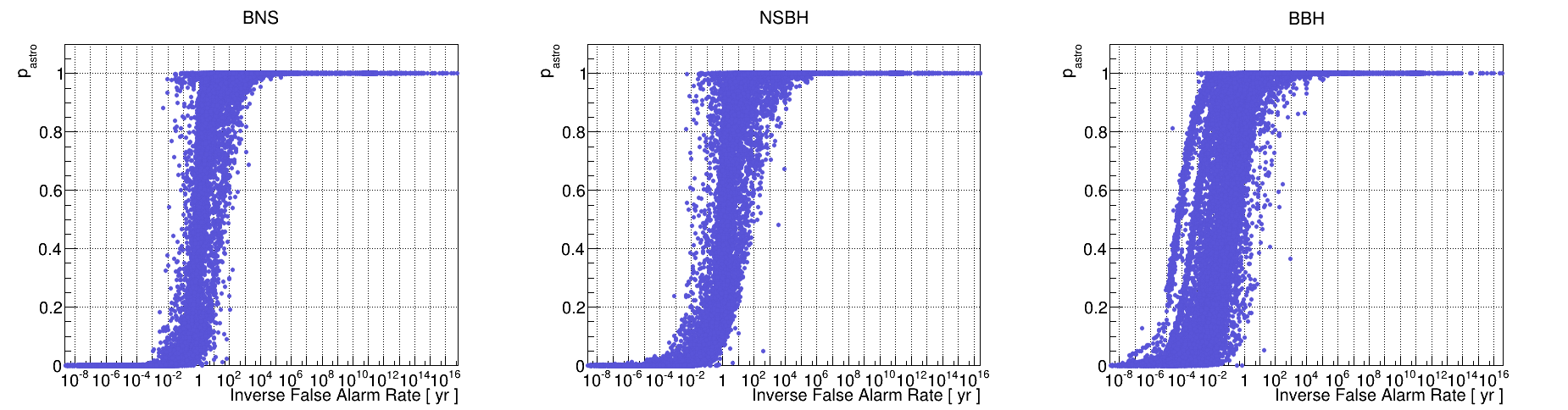}}
\caption{$p_{astro}$ as a function of the IFAR for the three types of injections in O3a data. \label{fig:pastroifarinjectionso3a}}
\end{figure*}

To further illustrate the parameterization, we show in figure~\ref{fig:pastroifarinjectionso3a} the $p_{astro}$ values attributed to injections versus their IFAR.
As expected, $p_{astro}\sim 0$ corresponds to low IFAR values, whereas $p_{astro}\sim 1$ corresponds to high IFAR values.
However, since the parameter space is split into $165$ bins to compute the foreground and the background distributions, we expect several $p_{astro}$ trends as a function of the IFAR computed over three regions of the parameter space. 
The width of the $p_{astro}$ transition from $0$ to $1$ is thus maximized in the BBH region of the parameter space, as the foreground and background variations are higher than for NSBH and BNS.
From the background perspective, the template bank has large variations of density which are not precisely taken into account to compute the IFAR, especially for high-mass sytems.
It results in a population of high-mass BBH assigned $p_{astro}> 0.5$ with $\textsc{IFAR}\sim10^{-4}~yr$ because the background used for $p_{astro}$ is estimated with only a few templates.
This comment is similar as in the previous analysis of candidate events from \textsc{GWTC-2.1} in figure~\ref{fig:O3apastroifar} where, for instance, $\textsc{GW}190916\_200658$ has an $\textsc{IFAR}\sim 1.4\times 10^{-4}~yr$ and a $p_{astro}\sim 0.66$.

During future observing runs, the parameterization would need to be updated as the number of events for each astrophysical category will increase and improve our knowledge of CBC merger rates.
For a sufficient number of detections, the parameterization should stabilize and the uncertainty on $p_{astro}$ should decrease.

\section{Classification} \label{classification}
Using $\sim$40000 injections of each type of sources and the $p_{astro}$ parameterization, we first assess the performance of our source classification, by looking at the class $\alpha$ with the highest $p_{\alpha}$.

\begin{figure} [h]
\centerline{\includegraphics[width=0.5\textwidth]{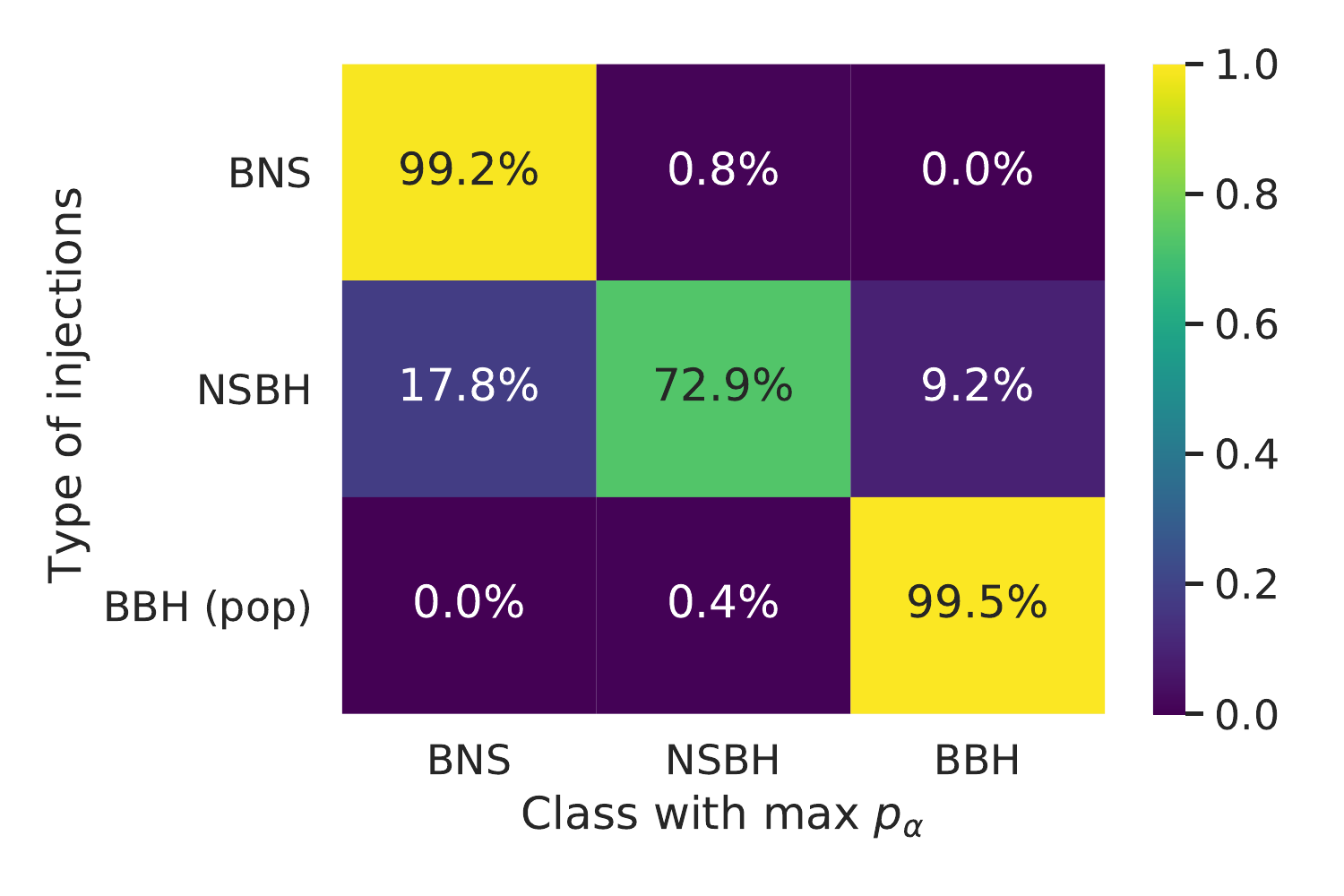}}
\caption{Confusion matrix of simulated GW signals in O3a data. The parameters used for the injections are described in table~\ref{tab:param_injections}.\label{fig:confmatrixo3a}}
\end{figure}

In figure~\ref{fig:confmatrixo3a}, the confusion matrix shows that the source classification is largely reliable for BNS and BBH injections, whereas a significant fraction of NSBH systems are categorized as BNS or BBH.
This is expected due to the poor measurement of the mass ratio $q$, which results in an overlap between some of the NSBH bins and the two other categories.

\begin{figure} [h]
\centerline{\includegraphics[width=0.55\textwidth]{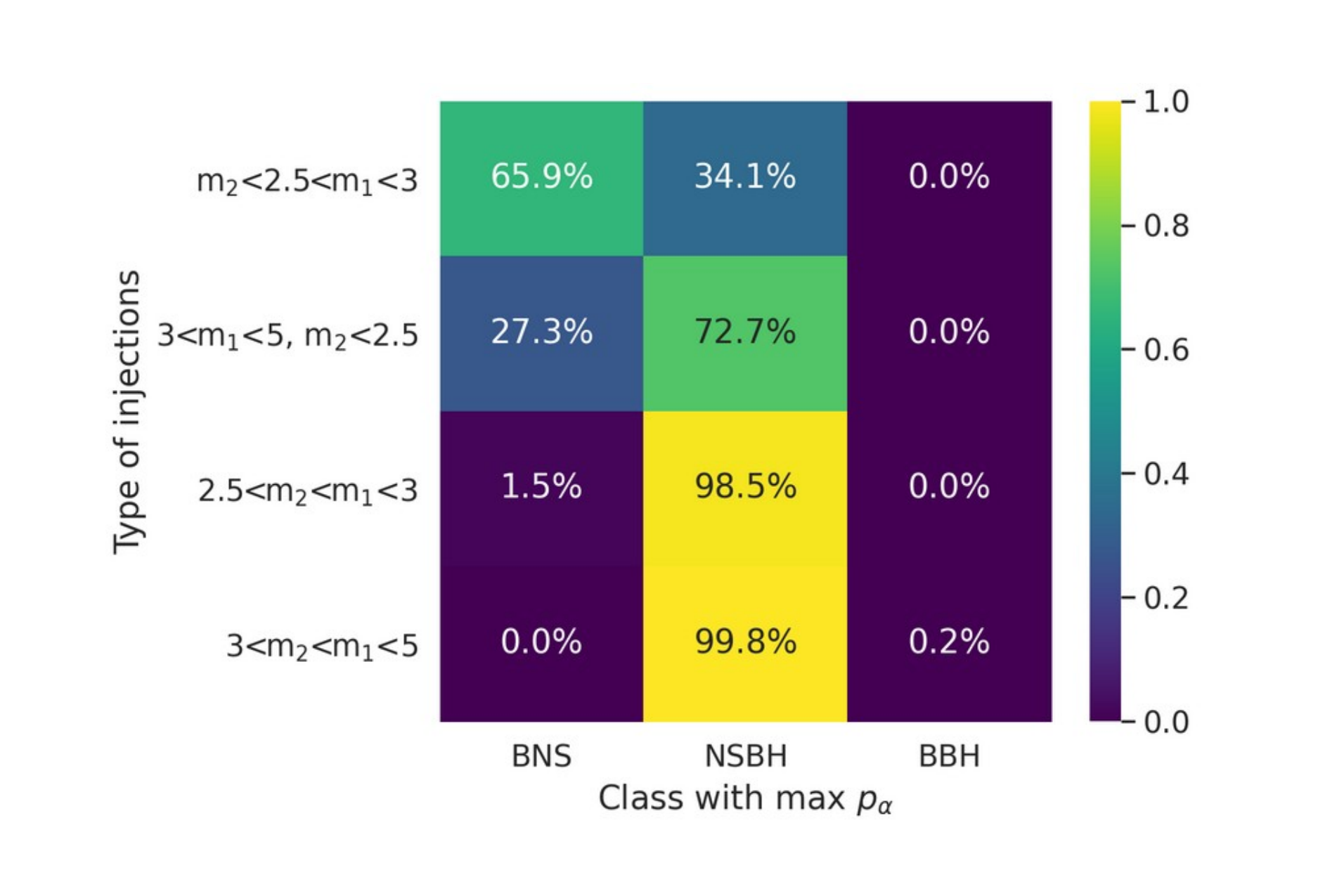}}
\caption{Confusion matrix of simulated GW signals for sources with at least one component mass between $2.5$ and $5~M_{\odot}$ in O3a data.\label{fig:confmatrixmassgapo3a}}
\end{figure}

In GWTC-2.1 for MBTA, as in this paper, a value of $2.5~M_{\odot}$ has been chosen for the NS maximum mass, with a gap to $5~M_{\odot}$ for BBH (but no gap for NSBH).
In figure~\ref{fig:confmatrixmassgapo3a}, we show how sources that would be present in this mass gap are categorized.
One can notice that none of the injections with masses below $3~M_{\odot}$ (choice made by other pipelines in GWTC-2.1) is classified as BBH.
Hence, our population models do not prevent us from labeling sources with masses below $3~M_{\odot}$ as BNS or NSBH, with the expected benefit of providing a cleaner source-class assessment.

\section{Sensitivity of $p_{astro}$ to assumptions} \label{uncertainties}
The $p_{astro}$ values are subject to various types of assumptions.
In general, triggers categorized as noise ($p_{astro} \sim 0$) and candidate events identified as real GW signals ($p_{astro} \sim 1$) are expected to be robust under different assumptions to compute $p_{astro}$.
However, candidate events with $0.1~\leq~p_{astro}~\leq~0.9$ are more likely to fluctuate according to different choices made to compute $p_{astro}$ and these fluctuations are maximized around $x_{50\%}$.
Possible sources of uncertainties on $p_{astro}$ considered in what follows are the ranking statistic fluctuations, the background fluctuations over the observing time, the statistical uncertainty in the number of expected events in the astrophysical categories, the way the template bank is divided into bins, or the choice of population models to build the foreground.
We study the effects of these sources by propagating them into uncertainties on $p_{astro}$ by using the parameterization in the case of HL-Von coincidences that would be found as BBH candidates (to be consistent with most of the MBTA detections in GWTC-2.1).
Then, we study the impact of the choice of the BBH and BNS population models on $p_{astro}$ using injections with $\mathrm{FAR}~<~1/hour$.
Finally, we quantify the impact of including or not the relative time spent into different detector network configurations in the foreground normalization.

\subsection{Impact of the combined ranking statistic fluctuations}

Regardless of the statistical uncertainty of $\pm1$ (for stationary noise) on the individual detector SNR, we focus on the systematic uncertainty on $x$ due to the discrete nature of the template bank. 
Since the latter uses a SNR minimal match of $97\%$ \cite{MBTA:2021, Brown:2012}, the expected maximum error on the SNR of an event is $\sim3\%$.
The related uncertainty on the SNR is taken as $3\%/\sqrt{12}\sim~0.9\%$ for a uniform distribution and we assume that the uncertainty on $x$ is the same as the uncertainty on the SNR around $x_{50\%}$.

We add fluctuations of $\pm 0.9\%$ on $x_{50\%}$ in the bins of the parameter space and compute a weighted mean of $p_{astro}$ variations using the distributions of expected BBH sources.
It gives an uncertainty of $\pm 0.08$ at $p_{astro}~=~0.5$.

This uncertainty could be reduced, for instance, to $\pm0.03$ if the SNR minimal match was increased from $97\%$ to $99\%$.

\subsection{Impact of the background fluctuations}

The $p_{astro}$ values are sensitive to choices in the background estimation.
In MBTA, the background for $p_{astro}$ was estimated over a few months of coincident analysis time to better sample the non-Gaussian fluctuations and to better populate the tail of the background distribution.
In order to get a quantitative estimate of the background fluctuations at $x_{50\%}$ and thus obtain the related systematic uncertainty on $p_{astro}$, we compute the distribution of background fluctuations at $\langle x_{50\%}\rangle=8.43$ (weighted mean value of $x_{50\%}$ using the distributions of expected BBH sources) over O3a using background estimated over periods of about one week.
The standard deviation of the mean background value is $20\%$.
This translates into a systematic uncertainty of $\pm0.05$ at $p_{astro}=0.5$.

Although this uncertainty is not the dominant one, it would be challenging to reduce it since the background fluctuations are driven by the state of the detectors which is not stationary during observing periods.

\subsection{Impact of the binning of the parameter space}

\begin{figure} [h]
\centerline{\includegraphics[width=0.5\textwidth]{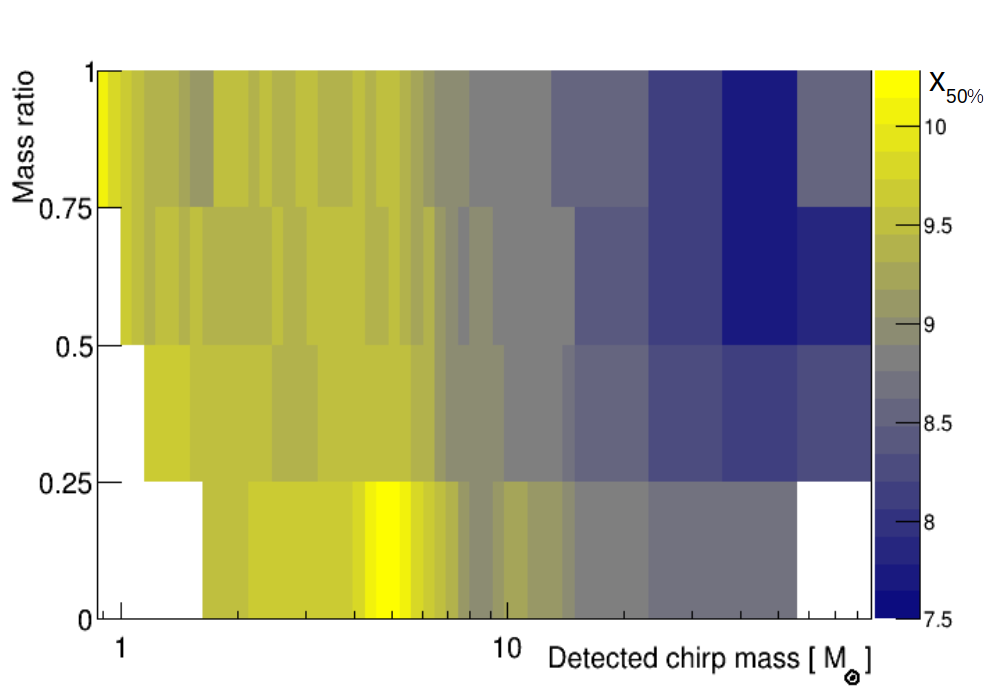}}
\caption{2D-histogram of $x_{50\%}$ for HL-Von coincidences in the ``$\mathcal{M}-q$" space. \label{fig:pastro50pcvar}}
\end{figure}

The division of the parameter space into bins, as described in section \ref{sec:binning}, is somewhat arbitrary and induces an uncertainty on $p_{astro}$ due to the discrete nature of the grid which does not allow a perfect tracking of the astrophysical population distributions, nor of the background distributions.
This means that the value of $p_{astro}$ for candidates whose parameters are close to bin boundaries is sensitive to changes in bin size.
We thus want to translate this sensitivity into an uncertainty, taking HL-Von BBH as an example.
Figure \ref{fig:pastro50pcvar} shows the $x_{50\%}$ variations over the ``$\mathcal{M}-q$" space for these coincidences.
The estimation is done considering the dispersion of $x_{50\%}$ between neighbouring bins.
First, the local variations of $p_{astro}$ values around a bin are weighted according to the expected BBH population inside the different bins.
Then a weighted mean of $p_{astro}$ variations over the entire parameter space is computed.
The weighted mean value is $0.23$ at $x_{50\%}$ while the median value of the variations is $0.15$.
The corresponding values along the $\mathcal{M}$ dimension are $0.19$ for the weighted mean and $0.10$ for the median; and along $q$, the weighted mean is $0.20$ and the median is $0.18$. Overall, $x_{50\%}$ varies smoothly over the parameter space.
However, as shown in figure \ref{fig:pastro50pcvar}, there are a few high-mass bins for which the variations increase rapidly.
This results in an average dispersion of $p_{astro}$ almost a factor of $2$ higher than the median dispersion considering chirp mass bins.
If one considers only mass ratio bins, the averaged and median values are much closer, which reflects the coarse division in that dimension.
As the average dispersion of $x_{50\%}$ over the parameter space is skewed by a few high-mass bins, we choose the median value of $0.15$ to estimate the systematic uncertainty at $p_{astro}\sim0.5$.

These results show the limit of the choice of the binning we made in some parts of the parameter space where the width of the bins could be further tuned to be better suited in the future.

\subsection{Impact of the expected number of astrophysical events}

The statistical uncertainty on the expected number of astrophysical events was marginalized over the posterior distribution of counts in the $p_{astro}$ computation for the MBTA candidate events in GWTC-2.1.
However, for the parameterization of $p_{astro}$, we use the mean values of the posterior distributions of astrophysical counts for the three categories.
We can thus estimate the statistical uncertainty on $p_{astro}$ values from the width of the distributions.
Since MBTA only found BBH candidate events during O3a with $p_{astro}~\sim~p_{BBH}$, we focus on the impact of the BBH counts on $p_{astro}$ values and especially around $x_{50\%}$.
We consider the $68\%$ confidence interval (CI) of the posterior distribution of BBH counts to choose the new values of counts and compute $p_{astro}$.
Let's call $\Lambda^{max}_{BBH} = 38.7$ (resp. $\Lambda^{min}_{BBH} = 26.3$) the upper (resp. lower) bound of the $68\%$ CI and let's consider $p_{astro}=p_{BBH}$.
Using equation \ref{eq:basicpastro} to compute $p_{astro}$ with the mean value of the BBH counts and $p^{max(min)}_{astro}$ with $\Lambda^{max(min)}_{BBH}$, one can compute a statistical uncertainty of $\pm0.05$ at $p_{astro}=0.5$. 

Future observing runs with better sensitivities of the detectors will help reducing this uncertainty as the number of GW sources detected will increase. 

\subsection{Impact of population models}
In this section, we focus on the impact of population models by testing the variation in $p_{astro}$ induced by different astrophysical models for BBH and BNS \cite{O3aAstroDist:2021, Alsing:2018, Chatziioannou:2020, Landry:2021}.
This is technically performed by reweighting the BBH and BNS injection sets with different populations in the source frame, which only affects the foreground distributions.
We do not study the NSBH case but the uncertainties derived for BBH and BNS can be considered as lower bounds for the NSBH case.

\paragraph{The BBH case}
~\\\\
As stated in section \ref{foreg}, for BBH we use the \textsc{Power Law + Peak} model without any redshift evolution of the merger rate, motivated by our current knowledge on BBH populations from the LIGO-Virgo data \cite{O3aAstroDist:2021}.
An alternative model considered in \cite{O3aAstroDist:2021} is the \textsc{Truncated} model, which is a simple power-law with a low- and a high-mass cutoffs.
This is the model which differs the most from the \textsc{Power~Law~+~Peak} \cite{O3aAstroDist:2021} and it is thus a good way to assess the worst case scenario.
Following \cite{O3aAstroDist:2021}, we use the subsequent parameters for the \textsc{Truncated} model: $m_{min} = 5~M_{\odot}$, $m_{max}~=~78.5~M_{\odot}$, $\alpha=-2.35$, $\beta = 1$ and $\kappa = 0$.

\begin{figure} [h]
\centerline{\includegraphics[width=0.5\textwidth]{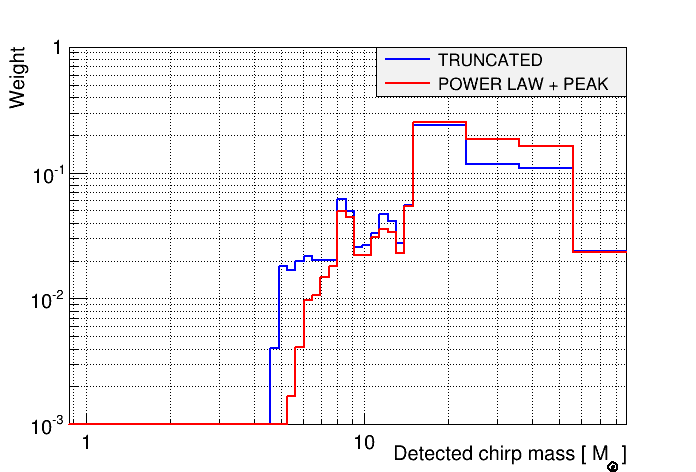}}
\caption{Normalized distributions of the detected chirp mass for the recovered BBH injections under two different population models. \label{fig:BBHTruncPLP}}
\end{figure}

\begin{figure*}[h]
  \centering
  \subfloat[]{\includegraphics[width=0.45\textwidth]{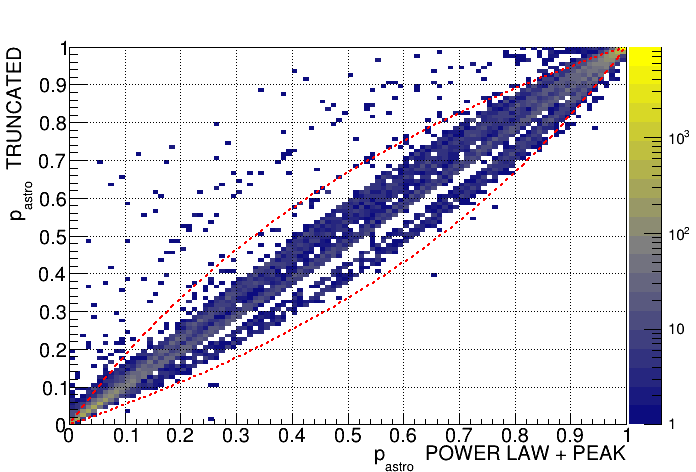}\label{fig:BBHscatter}}
  \hfill
  \subfloat[]{\includegraphics[width=0.45\textwidth]{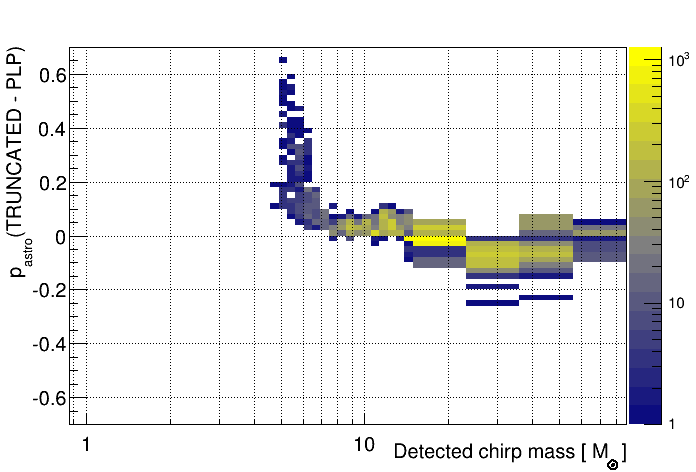}\label{fig:BBHdiff}}
  \caption{(a) 2D-histogram of $p_{astro}$ values computed using the \textsc{Truncated} model as a function of the $p_{astro}$ values computed with the \textsc{Power Law + Peak} model for the BBH injections with a $\textsc{FAR}<1/hour$. The red dashed lines serve as an example to show the effect of a constant factor 2 between the expected foreground of the two models. (b) 2D-histogram of the difference in $p_{astro}$ values between the \textsc{Truncated} and the \textsc{Power Law + Peak} models as a function of the detected chirp mass. Only BBH injections with $0.1 \leq p_{astro} \leq 0.9$ using either model are shown.}
\end{figure*}

Figure \ref{fig:BBHTruncPLP} shows the normalized distribution of the detected chirp mass for the BBH injections using the fiducial \textsc{Power~Law~+~Peak} model and the \textsc{Truncated} models.
One can see that the latter extends to lower chirp masses.
We then compute a new parameterization of $p_{astro}$ which is used on BBH injections.
In figure~\ref{fig:BBHscatter} we show a two-dimensional histogram of $p_{astro}$ values computed with both models.
As expected, most of the triggers with $p_{astro}$ close to $0$ or $1$ have similar values under one model or the other.
However, for $0.1 \leq p_{astro}~\leq~0.9$ a significant fraction of injections encounters fluctuations in their $p_{astro}$ with a discrepancy typically within $\pm0.1$ around $x_{50\%}$. One can also notice a small fraction of injections for which $p_{astro}$ differs significantly (with differences up to $0.6$) from one model to the other.
This difference can be seen in figure~\ref{fig:BBHdiff} as a function of the detected chirp mass where only injections with $0.1 \leq p_{astro} \leq 0.9$ have been used.

The median $p_{astro}$ difference of these injections is
$-0.002^{+0.085}_{-0.089}$ with $90\%$ CI\footnote{We choose the $90\%$ CI here to be sure to get a significant fraction of injections with a $p_{astro}$ around 0.5.}. The systematic uncertainty level is thus expected to be $\leq 0.1$ for most of the $p_{astro}$ values between $0.1$ and $0.9$ but some of the differences are much larger than the quoted CI in specific parts of the parameter space.

The largest differences in $p_{astro}$ occur where the foreground distributions of the two models differ most significantly, i.e. at low chirp mass.
Then, in the region of chirp masses of $30~M_{\odot} - 40~M_{\odot}$ the \textsc{Power Law + Peak} model gives higher $p_{astro}$ values than the other model mainly due to the Gaussian peak in the primary mass distribution.
Future observations will help reducing the systematic uncertainty coming from the BBH population model.

\paragraph{The BNS case}
~\\\\
The BNS injections used in the $p_{astro}$ calculation have been distributed using a uniform mass prior which was first motivated by the poor knowledge of the BNS population and which is also consistent with a NS population study performed on GW events only \cite{Landry:2021}.
However, the number of NS observed in GW events is small (two BNS \cite{GW170817:2017, GW190425:2020}, two NSBH \cite{NSBHdiscovery:2021} and one GW event with uncertain CBC category \cite{GW190814:2020}) and, for instance, the \textsc{Bimodal} model inferred from the galactic distribution of NS is not ruled out.
Hence, we might assume this model for the BNS population, as described in \cite{Alsing:2018, Chatziioannou:2020}, to compute $p_{astro}$.
More specifically, we use the the following parameters:
$m_{min} = 1~M_{\odot}$,  $\alpha = 0.65$, $\beta = 3$, $\mu_1 = 1.34~M_{\odot}$, $\sigma_1 = 0.07~M_{\odot}$, $\mu_2 = 1.80~M_{\odot}$, $\sigma_2 = 0.21~M_{\odot}$ and $m_{max} = 2.22~M_{\odot}$ from \cite{Landry:2020}, a maximum NS mass compatible with the galactic NS and GW events.

\begin{figure} [h]
\centerline{\includegraphics[width=0.5\textwidth]{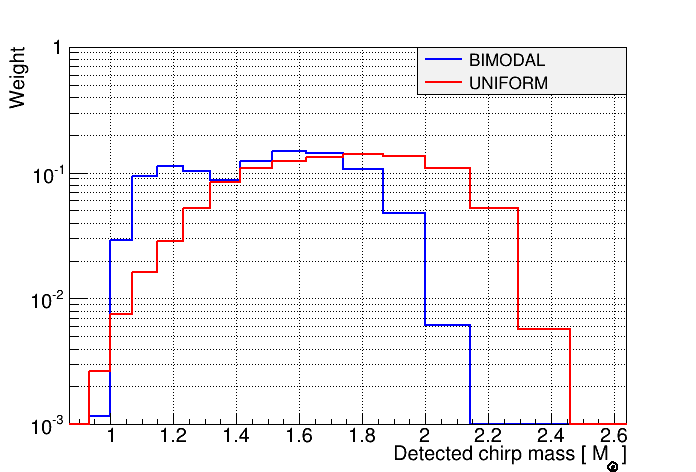}}
\caption{Normalized distributions of the detected chirp mass for the recovered BNS injections under two different population models.\label{fig:BNSGalUni}}
\end{figure}

Figure \ref{fig:BNSGalUni} shows the normalized distribution of the detected chirp mass for the BNS injections using the \textsc{Uniform} model and the \textsc{Bimodal} model.
The latter favors the low chirp mass region.

\begin{figure*}[h]
  \centering
  \subfloat[]{\includegraphics[width=0.45\textwidth]{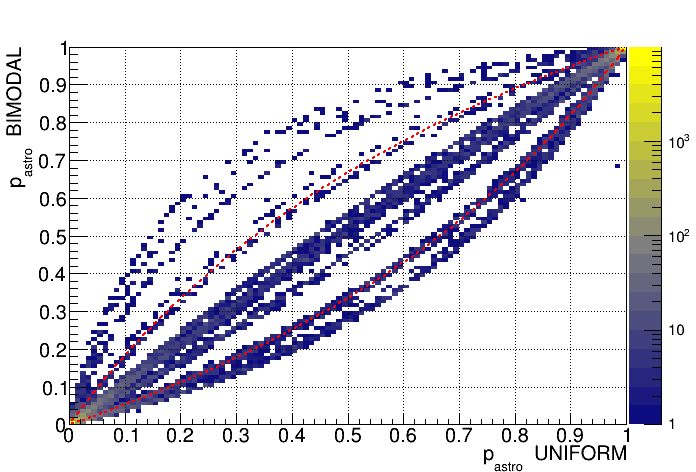}\label{fig:BNSscatter}}
  \hfill
  \subfloat[]{\includegraphics[width=0.45\textwidth]{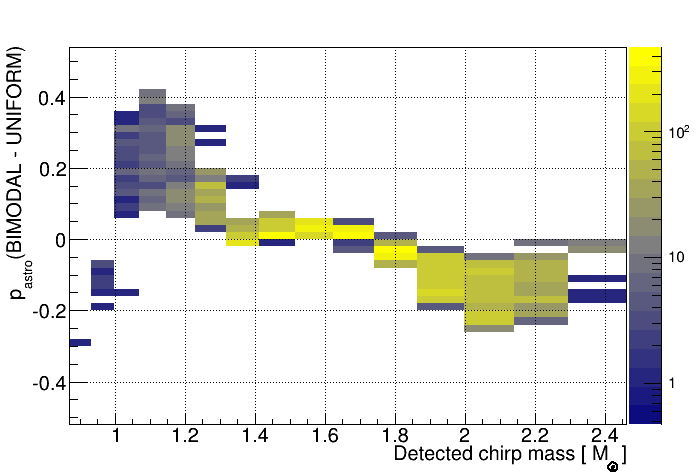}\label{fig:BNSdiff}}
  \caption{(a) 2D-histogram of $p_{astro}$ values computed using the \textsc{Bimodal} model as a function of the $p_{astro}$ values computed with the \textsc{Uniform} model for the BNS injections with a $\mathrm{FAR}~<~1/hour$. The red dashed lines serve as an example to show the effect of a constant factor 2 between the expected foreground of the two models. (b) 2D-histogram of the difference in $p_{astro}$ values between the \textsc{Bimodal} and the \textsc{Uniform} models as a function of the detected chirp mass for the BNS injections. Only BNS injections with $0.1 \leq p_{astro} \leq 0.9$ using either model are shown.}
\end{figure*}

As for the BBH injections, we construct a new set of parameterized $p_{astro}$ curves using the \textsc{Bimodal} model for BNS.
Figure~\ref{fig:BNSscatter} shows the two-dimensional histogram of BNS injections in the $p_{astro}$ plane for both models.
As in the case of the BBH injections, most of the BNS injections recovered with $p_{astro}\sim0$ or $1$ are almost unsensitive to population changes.
However, for injections with $0.1\leq p_{astro} \leq 0.9$ the discrepancy in $p_{astro}$ arising from the different populations is significantly higher (between 0.1 and 0.4) and grows toward $p_{astro}=0.5$.
The segmentation of the search parameter space also adds a $p_{astro}$ systematic uncertainty which can be assessed by the width of the empty regions between lines drawn by the parameterized curves.
The maximum gap in $p_{astro}$ values between two parameterized curves is $\sim 0.1$.
In the future, it may be worth considering reducing the uncertainty coming from the binning of the search parameter space by increasing the number of bins or by interpolating the parameterized curves to get a smooth transition from one bin to another.

To better see where most of the differences in $p_{astro}$ occur in the search parameter space, we show them as a function of the detected chirp mass in figure~\ref{fig:BNSdiff}, keeping only recovered injections with $0.1~\leq~p_{astro}~\leq~0.9$.
As expected, the differences follow closely the ones in the foreground distributions of both models seen in figure~\ref{fig:BNSGalUni}.
The median value of these differences with $90\%$ CI is $-0.002^{+0.156}_{-0.191}$. 
We thus expect the $p_{astro}$ systematic uncertainty level for BNS candidates to be $\leq 0.2$ depending on the assumed BNS population model.

Future observations of such systems will be essential in our understanding of their population and they will help getting more accurate $p_{astro}$ values. 

\subsection{Impact of the foreground network normalization}

\begin{figure} [h]
\centerline{\includegraphics[width=0.46\textwidth]{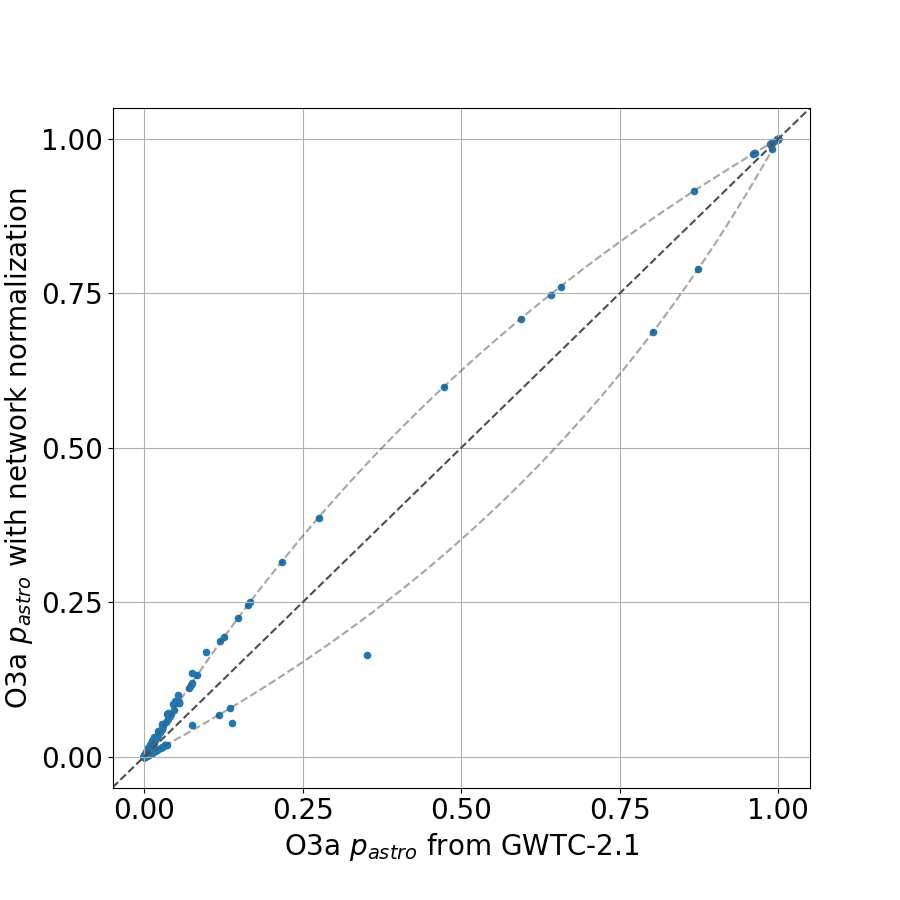}}
\caption{Scatter plot of the $p_{astro}$ values for the O3a triggers computed with the foreground network normalization versus the current $p_{astro}$ values from GWTC-2.1. The light gray dashed lines show the expected change in $p_{astro}$ for HL-Von (up) and HL (down) coincidence types.\label{fig:pastroscatternewforeg}}
\end{figure}

In section~\ref{foreg} we mentioned that the foreground weights and their normalization were computed for the O3 catalog papers using equal detector network configuration times.
However, the time spent in triple-detector configuration is $\sim55\%$ of the whole duration and the double-detector times account for $\sim15\%$ each.
To quantify the effect of this approximation, we compute $p_{astro}$ of O3a triggers using the normalized weights defined in equation~\ref{eq:newweights} and show the new values versus the current ones in figure~\ref{fig:pastroscatternewforeg}.

The $p_{astro}$ values for coincidences which occurred during triple detector time increase, whereas the ones during double detector times decrease.
The differences are of the same order of magnitude as the systematic uncertainties discussed in this section, with an overall trend to get slightly larger $p_{astro}$ values with the more accurate normalization (as there is more triple-detector time).
\section{Conclusion} \label{conclusions}
The method presented in this paper uses a multi-component analysis and population models that allow us to jointly perform a source classification and assign a probability of astrophysical origin to triggers \cite{Farr:2015, Kapadia:2020}.
One of the main features of the method is the division of the search parameter space to better capture the assumed population models of CBC sources and the use of a state-of the-art model for the population of BBH sources.
As a consequence, using a $p_{astro}$ threshold to select candidate events is of great interest compared to a selection on a $\mathrm{FAR}$ threshold since it may reveal BBH events in population-rich areas where the current MBTA $\mathrm{FAR}$ calculation is pessimistic.
We have also shown that $p_{astro}$ as a function of the combined ranking statistic squared can be parameterized in a simple manner based on the O3a data.
It allows for a fast estimation of $p_{astro}$ and source classification on large sets of injections and can be adapted to the online search for the next observing run O4 with a very low computational cost.

The $p_{astro}$ calculation is subject to several statistical and systematic uncertainties, which we have quantified.
The global systematic uncertainty on $p_{astro}$ has been estimated to $\mathcal{O}(0.1)$ in the region where $p_{astro}~\sim~0.5$ and the largest uncertainties lie in a few specific parts of the parameter space, where the population models are highly uncertain.
The uncertainties are large in the region where $p_{astro}\sim0.5$ and small for confident detections ($p_{astro}$ close to 1).
Using a $p_{astro}$ threshold of $0.5$ to select candidate events is a good way to not miss the interesting astrophysical events, at the price of a small contamination.
This is also a good choice since the number of candidates is minimal around that value, meaning the candidates list is rather stable against the uncertainties.

Future observations of GW signals will allow to refine some of the assumptions used in the $p_{astro}$ computation regarding population models and the number of expected astrophysical counts, and thus further reduce the subsequent $p_{astro}$ statistical and systematic uncertainties. 
Other sources of systematic uncertainties such as the SNR fluctuations or the binning of the parameter space can be reduced by increasing the template density inside the template bank and by dividing the parameter space into thinner bins.
\section*{Acknowledgments}
We express our gratitude to Gijs Nelemans and John T. Whelan, who have dedicated time and energy to perform internal LIGO-Virgo review of the MBTA pipeline. We thank our LIGO-Virgo collaborators from the CBC and low-latency groups for constructive comments. This analysis exploits the resources of the computing facility at the EGO-Virgo site, and of the Computing Center of the Institut National de Physique Nucl\'e{}aire et Physique des Particules (CC-IN2P3/CNRS).
LIGO Laboratory and Advanced LIGO are funded by the United States National Science Foundation (NSF) as well as the Science and Technology Facilities Council (STFC) of the United Kingdom, the Max-Planck-Society (MPS), and the State of Niedersachsen/Germany for support of the construction of Advanced LIGO and construction and operation of the GEO600 detector. Additional support for Advanced LIGO was provided by the Australian Research Council. Virgo is funded, through the European Gravitational Observatory (EGO), by the French Centre National de la Recherche Scientifique (CNRS), the Italian Istituto Nazionale di Fisica Nucleare (INFN) and the Dutch Nikhef, with contributions by institutions from Belgium, Germany, Greece, Hungary, Ireland, Japan, Monaco, Poland, Portugal, Spain.

\end{document}